%% file: root.tex
\documentclass[letterpaper, 10 pt, conference]{ieeeconf}  
\IEEEoverridecommandlockouts                              

\usepackage{hyperref}
\usepackage{amsmath}
\usepackage{graphicx}
\usepackage{pifont,xcolor} 
\usepackage{multirow}
\usepackage{stfloats}
\usepackage{booktabs} 
\usepackage{placeins}
\usepackage{subcaption}
\usepackage{array}
\usepackage{xcolor}
\usepackage{hyperref}
\usepackage{booktabs}  
\usepackage{pifont}    
\usepackage{listings}
\usepackage{tikz}

\usepackage[ruled,linesnumbered]{algorithm2e}

\usetikzlibrary{shapes.geometric, arrows.meta, positioning, calc}
\definecolor{slightdarkgreen}{rgb}{0,0.7,0} 
\newcommand{\cmark}{\textcolor{slightdarkgreen}{\ding{51}}}%
\newcommand{\xmark}{\textcolor{red}{\ding{55}}}%

\definecolor{nyupurple}{RGB}{87, 6, 140}

\title{\Huge
BladeChain: A Blockchain-based Traceability System for AI-Driven Engine Blade Inspection
}

\author{
{Mahmoud Hafez*$^1$}\hspace{2mm}
{Eman Ouda$^1$}\hspace{2mm}
{M. A. Mohammed Eltoum$^1$}\hspace{2mm}
{Khaled Salah$^2$}\hspace{2mm}
{Yusra Abdulrahman$^1$,$^3$}\thanks{*Corresponding author: mahmoud.hafez@ku.ac.ae}\\[4pt]
Khalifa University of Science \& Technology\\[4pt]
\thanks{$^{1}$Aerospace Engineering Department, Khalifa University of Science \& Technology}
\thanks{$^{2}$ Computer and Information Engineering Department, Khalifa University of Science \& Technology}
\thanks{$^{3}$ Advanced Research and Innovation Center (ARIC), Khalifa University of Science \& Technology}
}

\date{\today} 

\begin{document}

\maketitle
\thispagestyle{empty}
\pagestyle{empty}

\begin{abstract}
Aircraft engine blade maintenance relies on inspection records shared across manufacturers, airlines, maintenance organizations, and regulators. Yet current systems are fragmented, difficult to audit, and vulnerable to tampering. This paper presents BladeChain, a blockchain-based system providing immutable traceability for blade inspections throughout the component life cycle. BladeChain is the first system to integrate multi-stakeholder endorsement, automated inspection scheduling, AI model provenance, and cryptographic evidence binding, delivering auditable maintenance traceability for aerospace deployments. Built on a four-stakeholder Hyperledger Fabric network (OEM, Airline, MRO, Regulator), BladeChain captures every life cycle event in a tamper-evident ledger. A chaincode-enforced state machine governs blade status transitions and automatically triggers inspections when configurable flight hour, cycle, or calendar thresholds are exceeded, eliminating manual scheduling errors. Inspection artifacts are stored off-chain in IPFS and linked to on-chain records via SHA-256 hashes, with each inspection record capturing the AI model name and version used for defect detection. This enables regulators to audit both what defects were found and how they were found. The detection module is pluggable, allowing organizations to adopt or upgrade inspection models without modifying the ledger or workflows. We built a prototype and evaluated it on workloads of up to 100 blades, demonstrating 100\% life cycle completion with consistent throughput of 26 operations per minute. A centralized SQL baseline quantifies the consensus overhead and highlights the security trade-off. Security validation confirms tamper detection within 17~ms through hash verification.

\end{abstract}


\input{intro}
\input{related}

\input{system}
\input{implementation}

\input{eval}
\input{conc}
\input{ack}

\bibliographystyle{IEEEtran} 
\bibliography{root}

\end{document}

%% file: intro.tex
\section{Introduction}
\label{sec: introduction}

In the aviation industry, engine blades are among the most safety-critical serialized components~\cite{Qi2022fatiguereliability}. Their structural integrity governs engine efficiency, fuel burn, and, above all, flight safety~\cite{AbdullahAli2014bladelife}. Across a multi-year life cycle, each blade accumulates a chain of high-stakes events: manufacture and release, in-service operation, scheduled and unscheduled inspections, repairs and approvals, and ultimately removal from service~\cite{iata2020llp, Ho2021blockchainaircrafttrace}. Effective maintenance and inspection are therefore indispensable~\cite{PEREZGONZALEZ2005EASAreview}. Yet in practice, much of the associated evidence (images, annotations, findings, approvals) remains fragmented across paper forms, spreadsheets, and organization-specific databases, making it difficult to reconcile~\cite{FAA2012docchallenge, Agbaje2024AIAArecords, Shanmugam2015humanfactors}. These traditional information systems are often characterized by centralized storage, information lag, and high error rates, which act as a bottleneck for revenue and safety improvements~\cite{Xu2025a}. This fragmentation slows audits, weakens accountability, and increases the risk of incomplete or disputed maintenance histories~\cite{Mendonca2021documentation, Jensen2022airchain, aleshi2019blockchainmodel}.

The scale of this challenge is substantial: the global Maintenance, Repair, and Overhaul (MRO) market is projected to exceed \$125 billion by the early 2030s, with engine maintenance alone accounting for 46\% of civil aviation MRO spending~\cite{Nam2023mrokorea, HUANG2026_46}. The International Air Transport Association (IATA) emphasizes that quality traceability data throughout a part's life cycle is essential for inventory accuracy, reduced maintenance error, and effective decision-making, which are requirements that fragmented systems struggle to meet~\cite{iata2020llp, Ho2021blockchainaircrafttrace, KSalah2021BlockchainAerospaceDefense}. Regulatory frameworks such as EASA Part-145 mandate Occurrence Reporting Systems (ORS), yet many implementations function as passive repositories rather than comprehensive systems capable of enforcing traceability across organizational boundaries~\cite{PEREZGONZALEZ2005EASAreview}. Poorly documented parts have contributed to aviation incidents, underscoring the need for verifiable record systems that span Original Equipment Manufacturers (OEMs), airlines, MROs, and regulators~\cite{Jensen2022airchain, aleshi2019blockchainmodel}.

In parallel, AI/ML-based inspection for engine blades has advanced significantly, including borescope image analysis, crack detection, and surface-anomaly segmentation~\cite{yusra2023aerobladedefectreview}. These approaches reduce manual burden and improve consistency relative to purely visual inspection. However, most deployments remain point solutions: findings are confined to local Quality Assurance (QA) tools without cross-organization provenance, model-version accountability, or cryptographic bindings to the artifacts reviewed~\cite{yusra2023ai-blockchain-aerospace}. Even when an AI model flags a defect, there is no verifiable proof linking that output to the specific image, model version, responsible organization, and subsequent approval decision. The result is an infrastructure gap: AI can detect defects, but current systems do not anchor those detections into a regulator-grade audit trail.

Adopting Industry 4.0 standards, such as the use of IoT, Big Data, and Cloud Computing, is now seen as vital for improving supply chain integration and making more accurate operational decisions \cite{Oliveira-Dias2022}. Specifically, the transition to blockchain-enabled quality inspection can help supply chain members identify component problems earlier and approach ''zero-defect'' manufacturing \cite{Zhang2025}.While recent work has explored blockchain for aviation traceability, notably Hyperledger Fabric for spare parts inventory~\cite{Ho2021blockchainaircrafttrace}, permissioned ledgers for maintenance records~\cite{Jensen2022airchain, aleshi2019blockchainmodel}, and broader aerospace applications~\cite{KSalah2021BlockchainAviation, KSalah2021BlockchainAerospaceDefense, efthymiou2022exploratoryresearch}, these efforts focus on parts provenance rather than inspection-artifact binding, and none integrate AI-based defect detection with cryptographic commitments within a unified life cycle framework~\cite{yusra2022ai-blockchain-review, yusra2023ai-blockchain-aerospace}.

A permissioned blockchain is well-suited to this problem: multiple mutually distrusting organizations (OEMs, airlines, MROs, regulators) must jointly maintain authoritative records, no single party should unilaterally alter inspection evidence, and all state transitions require cryptographic proof of organizational endorsement~\cite{Wust_2018_DoWeNeedBlockChain, fabric}. Shared authority, tamper-evidence, and verifiable provenance are difficult to achieve with centralized databases or federated systems lacking consensus guarantees.

To address these gaps, we present \emph{BladeChain}, a blockchain-based framework for end-to-end traceability of aircraft engine blade inspections. BladeChain unifies a permissioned Hyperledger Fabric ledger governing blade life cycle events across OEM, airline, MRO, and regulator organizations; a pluggable AI-assisted inspection pipeline with off-chain artifact storage and on-chain cryptographic bindings; and an operational oracle layer that emits machine-verifiable traceability proofs. Our key contributions can be summarized as:

\begin{enumerate}
    \item We propose \textbf{BladeChain}, a blockchain-based traceability system specifically designed for AI-driven engine blade inspection, addressing the fragmented and non-auditable nature of current aviation maintenance records.

    \item We present \textbf{on-chain recording of AI model provenance}, where each inspection record captures the AI model name and version used for defect detection, enabling regulators to audit not only what defects were found but how they were found.

    \item We implement \textbf{automated inspection scheduling} via a chaincode-enforced state machine that triggers inspections based on configurable flight-hour, cycle, and calendar thresholds, eliminating manual tracking errors common in current maintenance workflows.
    
    \item We evaluate a working prototype on Hyperledger Fabric, demonstrating stable throughput and 100\% life cycle completion under realistic workloads, with a centralized SQL baseline quantifying the security-performance trade-off.
\end{enumerate}

The remainder of this paper is structured as follows. Section \ref{sec:related-work} reviews related work on blockchain applications for aircraft parts traceability, maintenance records, and MRO organizations, and positions the proposed BladeChain framework within existing research. Section \ref{sec:system-design} details the system design, including design requirements and decisions, system architecture, network topology, data model, and the associated security and threat model. Section \ref{sec:implementation} describes the implementation aspects, covering the development stack, chaincode implementation, API gateway and off-chain services, as well as the integration of defect detection. Section \ref{sec:evaluation} presents the experimental setup and evaluates the system through functional, performance, and security validations. Finally, Section \ref{sec:conclusion} concludes the paper by summarizing the key findings and outlining future research directions.

%% file: related.tex
\section{Related Work}
\label{sec:related-work}

Blockchain technology has attracted considerable attention in aviation for its potential to address longstanding challenges in parts traceability, maintenance record integrity, and multi-stakeholder coordination. This section reviews existing work on blockchain applications in aircraft maintenance and supply chain management, then positions BladeChain relative to these efforts.

\subsection{Blockchain for Aircraft Parts Traceability}

\begin{table*}[!htbp]
\centering
\caption{Comparison of blockchain systems for aviation maintenance and traceability.}
\label{tab:comparison}
\renewcommand{\arraystretch}{1.5}
\resizebox{\textwidth}{!}{%
\begin{tabular}{lccccccccc}
\toprule
\textbf{System} & \textbf{Year} & \textbf{Multi-Org} & \textbf{Multi-Org} & \textbf{Maintenance } & \textbf{Parts Life} & \textbf{AI} & \textbf{Off-Chain} & \textbf{Off-Chain} \\
 & & \textbf{Actors} & \textbf{Endorsement} & \textbf{Record Keeping} & \textbf{Cycle Mgmt} & \textbf{Integration} & \textbf{Record Storage} & \textbf{Integrity Proofs} \\
\midrule
Aleshi et al.~\cite{aleshi2019blockchainmodel} & 2019 & \cmark & \xmark & \cmark & \xmark & \xmark & \xmark & \xmark \\
Schyga et al.~\cite{Schyga2019MRO} & 2019 & \cmark & \cmark & \cmark & \cmark & \xmark & \xmark & \xmark \\
Ho et al.~\cite{Ho2021blockchainaircrafttrace} & 2021 & \cmark & \cmark & \cmark & \cmark & \xmark & \xmark & \xmark \\
AirChain~\cite{Jensen2022airchain} & 2022 & \cmark & \xmark & \cmark & \xmark & \xmark & \xmark & \xmark \\
Hawashin et al.~\cite{Hawashin2024NFTUAV} & 2024 & \cmark & \xmark & \cmark & \cmark & \xmark & \cmark & \xmark \\
Kabashkin~\cite{Kabashkin2024NFTframework} & 2024 & \cmark & \xmark & \cmark & \cmark & \xmark & \cmark & \xmark \\
Alqaryuti et al.~\cite{Alqaryuti2025hydraulic} & 2025 & \cmark & \xmark & \cmark & \xmark & \xmark & \cmark & \xmark \\
\midrule
\textbf{BladeChain (Ours)} & 2026 & \cmark & \cmark & \cmark & \cmark & \cmark & \cmark & \cmark \\
\bottomrule
\end{tabular}%
}
\end{table*}

Several studies have explored blockchain for tracking aircraft components across supply chain stakeholders. Ho et al.~\cite{Ho2021blockchainaircrafttrace} present a comprehensive Hyperledger Fabric implementation for aircraft spare parts management, achieving over 2,000 transactions per second with multi-organization channels connecting OEMs, MROs, airlines, and logistics providers. Their system supports private chaincode for inter-organizational data sharing, but focuses on inventory management rather than inspection evidence. Their design does not bind inspection artifacts or evidentiary media (e.g., images or NDT outputs) to ledger entries.

Joeaneke et al.~\cite{Joeaneke2024AerospaceBlockchain} provide quantitative validation of blockchain's impact on aerospace supply chains through structural equation modeling, demonstrating significant effects on traceability ($\beta = 0.40$, $p < 0.001$) and transparency ($\beta = 0.38$, $p < 0.001$). Their analysis confirms blockchain's value for part authentication and provenance tracking.

Recent work has shifted toward NFT-based authentication. Hawashin et al.~\cite{Hawashin2024NFTUAV} introduce composable NFTs for UAV parts traceability, where parent NFTs represent assembled UAVs and child NFTs represent individual components. This hierarchical model enables digital twin integration with life cycle tracking, using IPFS for off-chain metadata storage. Kabashkin~\cite{Kabashkin2024NFTframework} proposes a broader NFT-based framework for aviation component life cycle management spanning eight application areas, including parts tracking, maintenance records, certification, and quality assurance, and also utilizes IPFS for decentralized storage. However, both works focus primarily on manufacturing-stage provenance, ownership transfer, and certification events rather than in-service inspection workflows with multi-organizational endorsement requirements.

\subsection{Blockchain for Aircraft Maintenance Records}

AirChain~\cite{Jensen2022airchain} represents a mature maintenance record system, storing both Cabin Log Book (CLB) and Technical Log Book (TLB) data on Hyperledger Besu with smart contract-based data management. Developed with China Airlines, AirChain demonstrates practical viability but lacks multi-organization endorsement policies and off-chain storage for large artifacts such as images or inspection media.

Aleshi et al.~\cite{aleshi2019blockchainmodel} propose the Secure Aircraft Maintenance Records (SAMR) system using Hyperledger Sawtooth, addressing homeland security implications of maintenance fraud by preventing forgery of FAA personnel signatures. However, the system remains at a conceptual stage without implementation validation.

The most feature-complete recent work is the blockchain-driven framework for aircraft hydraulic systems by Alqaryuti et al.~\cite{Alqaryuti2025hydraulic}, combining Ethereum smart contracts, time oracles for automated maintenance scheduling, and IPFS for off-chain document storage. Their cost analysis shows \$0.89 average transaction cost, demonstrating economic viability. However, the system addresses preventive maintenance scheduling rather than inspection evidence binding, and does not integrate AI-assisted defect detection. While maintenance documents are stored off-chain and cryptographically anchored on-chain, the system does not provide artifact-level integrity proofs or support AI-assisted inspection traceability.Future iterations could leverage AI-driven predictive analytics to identify key emissions or failure drivers, validating these insights against immutable blockchain records to ensure auditable explanations for all variables \cite{Fatorachian2025}.

Kusumastuti and Hartono~\cite{Kusumastuti2024reliability} implement a Ganache-based prototype for aircraft reliability and maintenance records, while Kabashkin~\cite{Kabashkin2024iceberg} proposes the ``Iceberg Model'' integrating AI, blockchain, and data analytics for aircraft health monitoring. It is one of the few works combining AI with blockchain, though it remains conceptual.

\subsection{Blockchain in MRO Organizations}

Efthymiou et al.~\cite{efthymiou2022blockchain} provide a qualitative analysis of barriers to blockchain adoption in MRO organizations through a case study of Dublin Aerospace, identifying information-sharing reluctance, regulatory complexity, and implementation costs as key challenges. Their findings suggest that consortium blockchains are ideal for MRO due to permissioned participation requirements. 

Schyga et al.~\cite{Schyga2019MRO} developed a working Hyperledger Fabric prototype for MRO documentation with smart contracts for automated workflows, but it lacks integration with off-chain storage. Tapia et al.~\cite{Tapia2025DIMFabric} report one of the most detailed experimental evaluations of Hyperledger Fabric combined with decentralized identity and verifiable credentials in an aeronautical manufacturing context, demonstrating low overhead under industrial constraints. However, the focus is on manufacturing data provenance rather than aircraft maintenance or inspection workflows.


\subsection{Positioning of BladeChain}

Table~\ref{tab:comparison} highlights that existing blockchain-based systems in aviation address parts traceability, maintenance record storage, or workflow automation largely in isolation. Prior work has focused either on supply chain provenance and ownership transfer, on maintenance scheduling and documentation, or on conceptual integrations of AI and blockchain, but none provide an integrated mechanism for binding inspection artifacts, inspection actors, and inspection outcomes across organizational boundaries.

In particular, existing systems do not address AI model accountability in inspection workflows. While some approaches incorporate digital twins, analytics, or conceptual AI integration, they do not record which model version produced a given defect detection, nor do they cryptographically bind inference outputs to the original and annotated inspection artifacts reviewed by human inspectors. As a result, inspection outcomes cannot be independently re-validated or audited once evidence is stored off-chain.

BladeChain addresses these limitations by treating AI-assisted inspection as a first-class, auditable process. For each inspection event, the system records on-chain the model name and version, detected defect classes (corrosion, nick, dent, scratch), defect counts, inspector and organizational identities, and timestamps, alongside cryptographic commitments to the original and annotated inspection images via their IPFS CIDs and SHA-256 hashes. This design enables explicit, auditor-verifiable off-chain integrity proofs, allowing regulators or third parties to independently retrieve inspection artifacts, recompute hashes, and validate their integrity and provenance against on-chain records.

BladeChain is the first system to combine multi-organization endorsement, verifiable off-chain integrity proofs, and AI inspection accountability within a single blockchain-based framework for aviation inspection and maintenance.

%% file: system.tex
\section{System Design}
\label{sec:system-design}

This section presents the design of BladeChain, a blockchain-based framework for the traceability of aircraft engine blade inspections. We begin by establishing requirements derived from stakeholder needs, then describe our design decisions, system architecture, network topology, data model, and security considerations.

\begin{figure*}[!htbp]
  \centering
  \includegraphics[width=\textwidth]{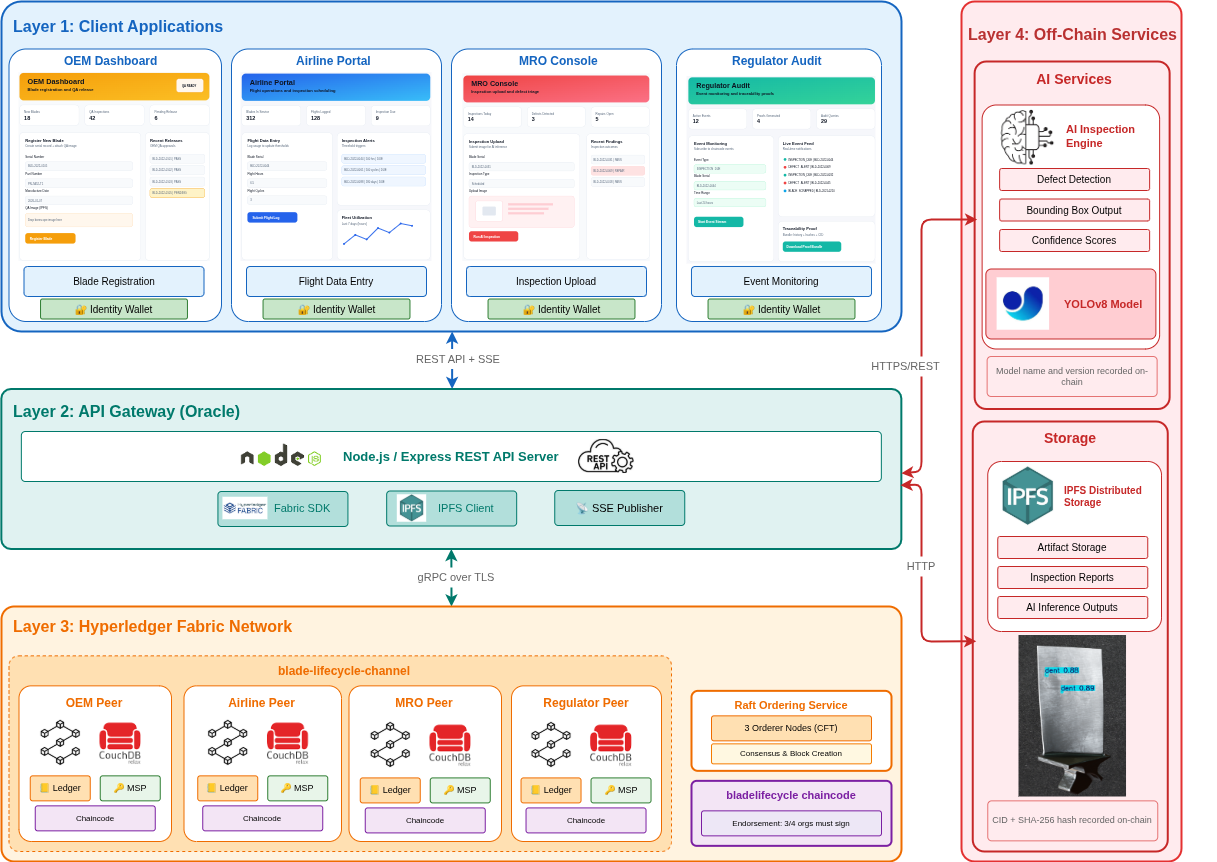}
  \caption{BladeChain architecture. Client applications communicate with the API gateway via REST/SSE. The gateway orchestrates interactions with the Hyperledger Fabric network, IPFS storage, and the pluggable AI inspection engine.}
  \label{fig:architecture}
\end{figure*}

\subsection{Design Requirements}
\label{subsec:requirements}

BladeChain must satisfy the requirements of four stakeholder classes (OEMs, airlines, MROs, and regulators), each with distinct concerns regarding blade lifecycle management.

\emph{\textbf{R1: Multi-Organization Traceability.}} All life cycle events shall be attributable to a specific organization and individual, with cryptographic non-repudiation.

\emph{\textbf{R2: Multi-Party Control.}} No single organization shall be able to unilaterally create, modify, or delete records.

\emph{\textbf{R3: Automated Inspection Scheduling.}} The system shall enforce inspection schedules based on configurable thresholds. When a blade exceeds any threshold, it shall automatically transition to an inspection-due state, preventing further flight operations until inspection is completed.

\emph{\textbf{R4: Evidence Binding.}} Inspection artifacts shall be cryptographically bound to on-chain records. This binding shall include: (i) tamper-evident commitments to artifact content, (ii) the AI model name and version used for defect detection, (iii) the inspector identity and organization, and (iv) timestamps for all events.

\emph{\textbf{R5: Auditability.}} Regulators shall be able to retrieve complete blade histories, verify artifact integrity, and receive real-time event notifications. The system shall support proof generation bundling on-chain records with hash revalidation.

\emph{\textbf{R6: Practical Scalability.}} The system shall support maintenance workloads typical of fleet operations. While high transaction throughput is not required, the system shall handle concurrent inspections across multiple blades without performance degradation.

\subsection{Design Decisions}
\label{subsec:decisions}

The requirements outlined above informed several architectural decisions regarding the selection of a blockchain platform, artifact storage, AI integration, and transaction endorsement.

We selected Hyperledger Fabric 2.5 as the underlying blockchain platform after evaluating alternatives, including public blockchains and other permissioned frameworks. Public blockchains were unsuitable for two reasons: their pseudonymous identity models conflict with regulatory requirements for accountable, auditable participants, and their consensus mechanisms introduce latency and cost (gas fees) unnecessary for maintenance workloads. Among permissioned alternatives, Fabric offered the most mature support for our requirements. Its Membership Service Provider (MSP) architecture provides X.509 certificate-based identity management aligned with enterprise PKI practices~\cite{Androulaki2018FabricEOV}. Endorsement policies enable fine-grained specification of which organizations must approve transactions, directly addressing the multi-organization traceability requirement. The execute--order--validate architecture decouples transaction execution from consensus, enabling parallel endorsement across organizations, while Raft-based ordering provides deterministic finality without the overhead of Byzantine fault--tolerant consensus mechanisms~\cite{fabric}.

For inspection artifacts, we adopted an off-chain storage strategy using IPFS instead of storing images directly on the ledger. Storing images on-chain would rapidly bloat the ledger, degrade network performance, and complicate peer synchronization. Instead, BladeChain stores artifacts in IPFS and records only the Content Identifier (CID) and SHA-256 hash on-chain. The CID enables content-addressed retrieval, while the hash provides an independent verification path: any auditor or stakeholder can independently retrieve the artifact and confirm integrity by recomputing the hash~\cite{Benet2014IPFS}. This approach mirrors patterns established in prior blockchain systems for document-heavy domains~\cite{Alqaryuti2025hydraulic, Kabashkin2024NFTframework}.

The defect detection component is designed as a pluggable module rather than a fixed system element, with all inference performed off-chain. While our prototype employs a YOLOv8-based detector, the architecture imposes no constraints on model architecture or vendor. The chaincode records model name and version alongside detection outputs, enabling post-hoc audit of which model produced which findings. This design acknowledges that detection technology will evolve and that different operators may prefer different solutions. BladeChain provides the traceability infrastructure without mandating specific AI implementations.

Finally, we configured Fabric's endorsement policy to require transaction signatures from at least three organizations configured as a logical Majority endorsement policy. This policy ensures that inspection records cannot be created through unilateral action; a majority of organizations must cryptographically endorse each transaction. The Regulator organization participates in the channel and may endorse transactions, but primarily serves an audit and oversight role.

\subsection{System Architecture}
\label{subsec:architecture}

BladeChain employs a four-layer architecture that separates client applications, the API gateway, the blockchain network, and off-chain services. Figure~\ref{fig:architecture} illustrates the overall system structure.

\emph{\textbf{Layer 1: Client Applications.}} The presentation layer comprises web-based dashboards tailored to each stakeholder role. Operators view blade status and initiate flight data updates; MRO technicians upload inspection images and review AI-generated annotations; regulators access audit interfaces and event streams. All clients interact with the system through the API gateway rather than directly with the blockchain, simplifying integration and enabling consistent access control.

\emph{\textbf{Layer 2: API Gateway (Oracle).}} The API gateway serves as the bridge between external clients and the blockchain network. Implemented as a Node.js service, it exposes RESTful endpoints for all lifecycle operations: blade manufacturing, activation, flight data recording, inspection submission, repair logging, and proof generation. The gateway also provides Server-Sent Events (SSE) for real-time event streaming, enabling clients to receive notifications when blade states change. Critically, the gateway orchestrates the inspection workflow: receiving uploaded images, invoking the AI detection engine, uploading artifacts to IPFS, computing SHA-256 hashes, and submitting transactions to the Fabric network.

\emph{\textbf{Layer 3: Hyperledger Fabric Network.}} The blockchain layer provides the immutable, multi-party ledger for blade life cycle records. The network comprises peers from each organization (with CouchDB state databases for rich queries), a Raft-based ordering service for transaction sequencing, and the \texttt{bladelifecycle} chaincode implementing the blade state machine and access control logic. All peer-to-peer communication uses gRPC over TLS.

\emph{\textbf{Layer 4: Off-Chain Services.}} Two off-chain components complement the blockchain. The AI inspection engine returns bounding boxes, defect labels produced by the model, and confidence scores. The IPFS daemon provides content-addressed storage for inspection artifacts, returning CIDs that are recorded on-chain. Both services are accessed by the API gateway during the inspection workflow.

\paragraph{Inspection Data Flow}

\begin{figure}[!htbp]
  \centering
  \includegraphics[width=\columnwidth]{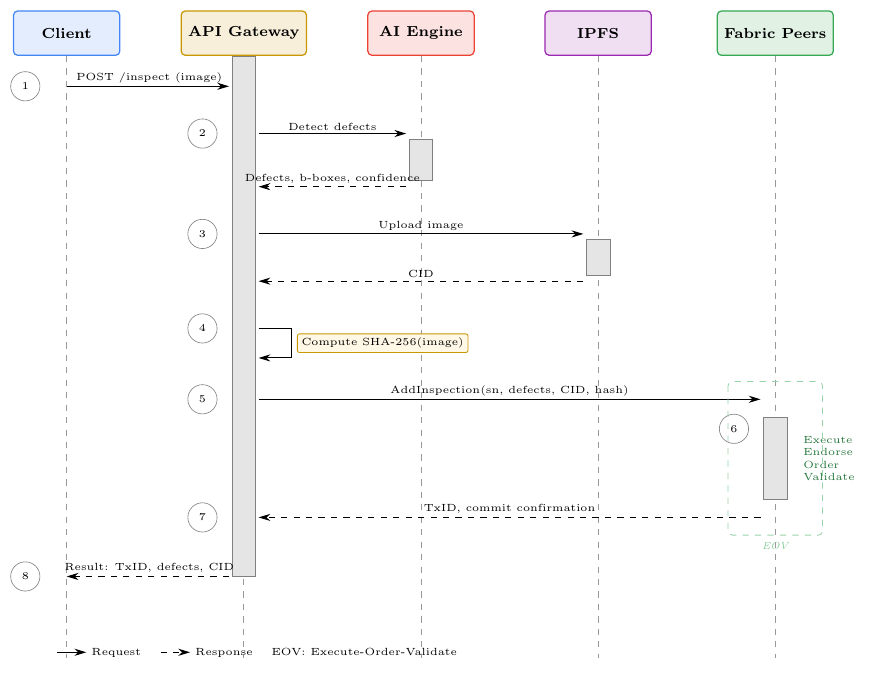}
  \caption{Sequence diagram for blade inspection operations. The API gateway orchestrates AI inference, IPFS storage, hash computation, and chaincode invocation before returning results to the client.}
  \label{fig:dataflow}
\end{figure}

Figure~\ref{fig:dataflow} illustrates the sequence of operations when an MRO technician submits a blade inspection.

The workflow proceeds as follows:
\begin{enumerate}
    \item The client uploads an inspection image to the API gateway via HTTP POST.
    \item The gateway forwards the image to the AI engine, which returns detected defects with bounding boxes and confidence scores.
    \item The gateway uploads the inspection artifact (annotated image when available) to IPFS, receiving a CID.
    \item The gateway computes the SHA-256 hash of the image content.
    \item The gateway invokes the \texttt{AddInspection} chaincode function with the blade serial number, defect list, model metadata, CID, and hash.
    \item Endorsing peers independently execute the chaincode and return signed endorsements.
    \item The gateway submits the endorsed transaction to the ordering service.
    \item Orderers sequence the transaction into a block and distribute it to all peers.
    \item Each peer validates the transaction against the endorsement policy and commits the state update to CouchDB.
    \item The gateway returns the transaction ID and inspection results to the client.
\end{enumerate}

\subsection{Network Topology}
\label{subsec:topology}

\begin{figure}[!htbp]
  \centering
  \includegraphics[width=\columnwidth]{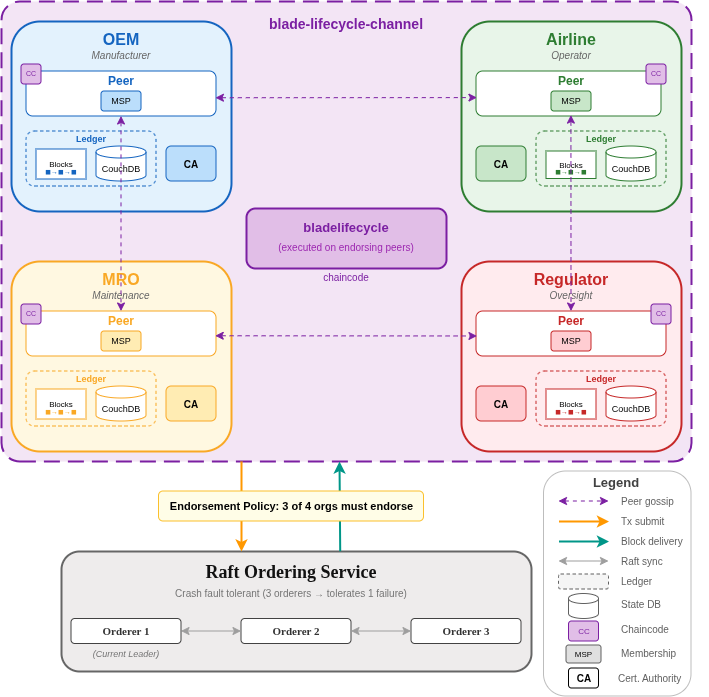}
  \caption{BladeChain network topology. Peers execute the chaincode and sign. Orderers sequence endorsed transactions into blocks. Peers validate and commit to state database.}
  \label{fig:topology}
\end{figure}

The BladeChain network comprises four organizations reflecting the aviation maintenance ecosystem. Figure~\ref{fig:topology} depicts the network structure.

\paragraph{Organizations} Four organizations participate in the network:
\begin{itemize}
    \item \textbf{OEM}: Manufactures blades, performs release inspections, provides technical authority.
    \item \textbf{Airline}: Operates aircraft, records flight data, owns blades during service.
    \item \textbf{MRO}: Performs inspections, executes repairs.
    \item \textbf{Regulator}: Audits records, verifies compliance.
\end{itemize}

\paragraph{Channel Configuration} All organizations share a single channel (\texttt{blade\-lifecycle-\-channel}) providing a unified view of blade records. The channel hosts the \texttt{bladelifecycle} chaincode, which implements the blade state machine and enforces access control based on caller identity.

\paragraph{Ordering Service} Transaction ordering uses a three-node Raft cluster providing crash fault tolerance (tolerating one node failure). Raft was selected over Byzantine fault-tolerant alternatives (e.g., BFT-SMaRt) because the permissioned network assumes non-Byzantine behavior among known ordering nodes, and Raft offers lower latency and simpler operations.

\paragraph{Peer Configuration} Each organization operates one peer node with an associated CouchDB instance for state storage. CouchDB enables rich queries against blade records (e.g., retrieving all blades in \texttt{INSPECTION\_DUE} state), supporting operational dashboards and audit interfaces.

\subsection{Data Model}
\label{subsec:datamodel}

The BladeChain data model captures the complete life cycle of aircraft engine blades through two primary constructs: a state machine governing blade status transitions, and structured records for inspections and defects.

\paragraph{Blade life cycle State Machine}

\begin{figure}[!htbp]
  \centering
  \includegraphics[width=\columnwidth]{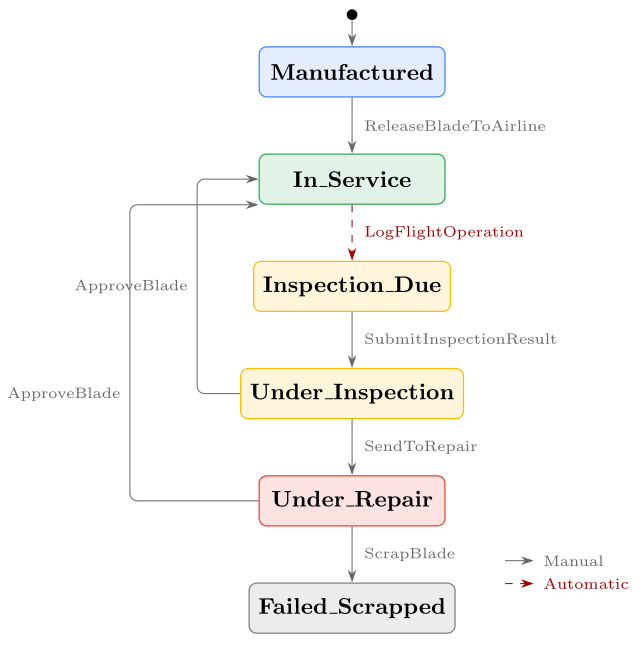}
  \caption{Blade life cycle stages and permitted transitions. Transitions are enforced by chaincode; the transition to \textsc{Inspection\_Due} occurs automatically when usage thresholds are exceeded.}
  \label{fig:statemachine}
\end{figure}

Each blade progresses through six states that reflect its operational status. A blade begins in the \textsc{Manufactured} state upon creation by the OEM and undergoes a quality assurance inspection before release. Upon release to an airline, the blade transitions to \textsc{In\_Service}, where it accumulates flight hours and cycles during normal operations. When any of the three configurable thresholds is exceeded, the chaincode automatically transitions the blade to \textsc{Inspection\_Due}, recording which threshold triggered the transition. When inspection results are submitted, the blade moves to \textsc{Under\_Inspection}. Based on the outcome, the blade may be approved to return to \textsc{In\_Service}, or sent to \textsc{Under\_Repair} if defects require remediation. Upon successful repair and approval, the blade returns to service. If damage is irreparable, the blade transitions to \textsc{Failed\_Scrapped} and is permanently removed from service. Figure~\ref{fig:statemachine} illustrates these states and the triggers for the transitions.

The automatic transition to \textsc{Inspection\_Due} is central to regulatory compliance. When \texttt{LogFlightOperation()} is invoked with new flight hours and cycles, the chaincode evaluates three threshold conditions and transitions the blade state if any condition is met. This design ensures that no blade can continue operating beyond its inspection interval without explicit inspection and approval.

\paragraph{On-Chain Records}
The blade record maintains both static attributes established at manufacture and dynamic fields updated throughout the life cycle. Table~\ref{tab:bladerecord} summarizes the key fields stored in CouchDB.


\vspace{3pt}

\begin{table}[!htbp]
\centering
\caption{Blade record fields stored on-chain.}
\label{tab:bladerecord}
\renewcommand{\arraystretch}{1.1}
\footnotesize
\begin{tabular}{ll}
\toprule
\textbf{Field} & \textbf{Description} \\
\midrule
\texttt{serialNumber} & Unique blade identifier (PK) \\
\texttt{currentState} & Current life cycle state \\
\texttt{currentOwner} & Operating airline (MSP ID) \\
\texttt{manufactureDate} & ISO 8601 manufacture timestamp \\
\texttt{totalFlightHours} & Cumulative flight hours \\
\texttt{totalFlightCycles} & Cumulative flight cycles \\
\texttt{lastFlightDate} & Most recent flight operation date\\
\texttt{hoursSinceInspection} & Hours since last inspection \\
\texttt{cyclesSinceInspection} & Cycles since last inspection \\
\texttt{daysSinceInspection} & Days since last inspection \\
\texttt{inspectionDueReason} & Threshold that trigger inspection \\
\texttt{nextInspectionDue} & Scheduled next inspection date \\
\texttt{lastInspectionDate} & Most recent inspection date \\
\texttt{totalInspections} & Count of completed inspections \\
\texttt{inspectionHistory} & Array of inspection event IDs \\
\texttt{createdAt} & Record creation timestamp \\
\texttt{updatedAt} & Last modification timestamp \\
\bottomrule
\end{tabular}
\vspace{5pt}
\end{table}

Each inspection generates an \texttt{InspectionEvent} record appended to the blade's history. Table~\ref{tab:inspectionevent} describes the inspection event structure.

\vspace{3pt}

\begin{table}[!htbp]
\centering
\caption{Inspection event fields recorded on-chain.}
\label{tab:inspectionevent}
\renewcommand{\arraystretch}{1.1}
\footnotesize
\begin{tabular}{ll}
\toprule
\textbf{Field} & \textbf{Description} \\
\midrule
\texttt{eventID} & Deterministic identifier \\
\texttt{serialNumber} & Associated blade serial number \\
\texttt{inspectionDate} & ISO 8601 inspection timestamp \\
\texttt{inspectionType} & Type \\
\texttt{aiModelName} & AI model name \\
\texttt{aiModelVersion} & AI model version \\
\texttt{detectedDefects} & Array of defect objects \\
\texttt{defectCount} & Total defects detected \\
\texttt{overallResult} & Outcome (pass / fail / critical) \\
\texttt{inspector} & Inspector identity (from MSP) \\
\texttt{organization} & Inspecting organization (MSP ID) \\
\texttt{imageIPFS} & IPFS CID of inspection image \\
\texttt{imageHash} & SHA-256 hash of image content \\
\texttt{flightHoursAtInspection} & Flight hours at inspection time \\
\texttt{flightCyclesAtInspection} & Flight cycles at inspection time \\
\texttt{notes} & Optional inspector notes \\
\texttt{timestamp} & Event timestamp \\
\bottomrule
\end{tabular}
\vspace{5pt}
\end{table}

Each defect in the \texttt{detectedDefects} array contains: \texttt{defectType} (classification from the AI model), \texttt{confidence} (0.0--1.0), and bounding box coordinates (\texttt{x1}, \texttt{y1}, \texttt{x2}, \texttt{y2}) localizing the defect in the image.

\paragraph{Manufacturing Inspection Normalization}

Manufacturing inspections (quality assurance at the OEM) follow a normalized format to ensure consistent record structure. The result defaults to PASS with an empty defect list, model name and version are set to ``N/A'' (no AI detection required for new blades), and the inspector and organization are recorded as ``OEM\_QA'' and ``OEM'' respectively. This normalization prevents null values in the data model while preserving the distinction between manufacturing and operational inspections.

\subsection{Security and Threat Model}
\label{subsec:security}

BladeChain's security model leverages Fabric-native mechanisms for authentication, authorization, and data integrity, while the threat model defines the assumptions underlying these guarantees.

\paragraph{Authentication and Authorization}

All network participants authenticate via X.509 certificates issued by organization-specific Certificate Authorities within Fabric's MSP framework. Transactions carry the submitter's certificate, enabling non-repudiable attribution of actions to specific identities. Inspector and organization fields in inspection records are derived from the caller's MSP certificate, ensuring that inspection provenance cannot be falsified by client applications.

At the network level, the channel endorsement policy requires a majority of the four organizations to sign transactions. At the chaincode level, authorization is enforced through state checks. For example, \texttt{LogFlightOperation} verifies the blade is in \textsc{In\_Service} state before accepting flight data.

\paragraph{Data Integrity}

Table~\ref{tab:integrity} summarizes the integrity mechanisms protecting on-chain and off-chain data.

\begin{table}[!htbp]
\vspace{5pt}
\centering
\caption{Data integrity mechanisms.}
\label{tab:integrity}
\renewcommand{\arraystretch}{1.5}
\begin{tabular}{p{2.2cm}p{5.5cm}}
\toprule
\textbf{Mechanism} & \textbf{Protection} \\
\midrule
Block hashing & Each block contains the hash of the previous block; tampering invalidates subsequent chain \\
Endorsement signatures & Committed transactions carry signatures from the majority of organizations \\
IPFS CID & Content-addressed storage; CID derived from content hash \\
SHA-256 binding & On-chain hash enables independent artifact verification \\
gRPC over TLS & Encrypted peer-to-peer and client-to-peer communication \\
\bottomrule
\end{tabular}
\vspace{5pt}
\end{table}

\paragraph{Auditability}

The chaincode provides a \texttt{GetBladeCompleteHistory} function returning the current blade record and all associated inspection events. The API gateway exposes a proof generation endpoint that bundles this history with recomputed artifact hashes for independent verification. Selective chaincode events notify subscribers of key state changes: \texttt{InspectionDueEvent} when thresholds trigger inspection requirements, \texttt{DefectDetectedEvent} and \texttt{OEMDefectAlert} when defects are found, and \texttt{BladeScrapedEvent} when blades are removed from service.

\paragraph{Threat Model}

Table~\ref{tab:threats} summarizes adversary capabilities and corresponding mitigations.

\begin{table}[!htbp]
\vspace{5pt}
\centering
\caption{Threat model: adversary capabilities and mitigations.}
\label{tab:threats}
\renewcommand{\arraystretch}{1.5}
\begin{tabular}{@{}p{2cm}p{5.9cm}@{}}
\toprule
\textbf{Threat} & \textbf{Mitigation} \\
\midrule
Invalid transaction & Rejected by majority endorsement policy \\
Single compromised peer & Cannot forge transactions without majority endorsement \\
Forged inspector identity & Inspector/org derived from MSP certificate, not client input \\
Network eavesdropper & TLS encryption on all channels \\
IPFS substitution & CID and SHA-256 verification detects modification \\
Raft node failure & Tolerates $<N/2$ crash faults (1 of 3 orderers) \\
\bottomrule
\end{tabular}
\vspace{5pt}
\end{table}

We assume an honest majority among endorsing organizations and that each organization's root CA issues certificates only to legitimate members. Out-of-scope threats include collusion among a majority of endorsers (requiring governance controls), adversarial attacks against the AI model (orthogonal ML security research), and denial-of-service (an infrastructure concern).

%% file: implementation.tex
\section{Implementation}
\label{sec:implementation}

This section describes the implementation of BladeChain, covering the development stack, chaincode logic, API layer, and defect detection integration.

\subsection{Development Stack}
\label{subsec:stack}

Table~\ref{tab:stack} summarizes the technologies and versions used in the BladeChain prototype.

\begin{table}[!htbp]
\vspace{5pt}
\centering
\caption{Implementation technology stack.}
\label{tab:stack}
\renewcommand{\arraystretch}{1.5}
\small

\begin{tabular}{lll}
\toprule
\textbf{Component} & \textbf{Technology} & \textbf{Version} \\
\midrule
Blockchain framework & Hyperledger Fabric & 2.5.14 \\
Consensus & Raft (etcdraft) & 3 orderers \\
Smart contract & Go & 1.21.13 \\
State database & Apache CouchDB & 3.3.2 \\
API gateway & Node.js + Express & 20.19.5 \\
Fabric SDK & fabric-network & 2.2.20 \\
Decentralized storage & IPFS (Kubo) & 0.24.0 \\
Defect detection & YOLOv8  & Custom\\
Containerization & Docker & 28.4.0 \\
\bottomrule
\end{tabular}
\vspace{5pt}
\end{table}

The network deployment consists of thirteen Docker containers: three orderers forming the Raft cluster, four peers (one per organization), four CouchDB instances for state storage, an IPFS daemon, and the API gateway. Cryptographic materials (X.509 certificates) are generated using Fabric's \texttt{cryptogen} tool for all organizations, peers, orderers, and user identities.

\subsection{Chaincode Implementation}
\label{subsec:chaincode}

The \texttt{bladelifecycle} chaincode, implemented in Go, encodes the blade state machine and exposes functions for each life cycle operation. 

\paragraph{Automatic Inspection Triggering}

A key safety feature is the automatic transition to \textsc{Inspection\_Due} when usage thresholds are exceeded. Algorithm~\ref{alg:flightop} presents the logic implemented in \texttt{LogFlightOperation}.

\begin{algorithm}[!htbp]
\caption{LogFlightOperation}
\label{alg:flightop}
\KwIn{serialNumber, flightHours, flightCycles, flightDate}
\KwOut{Updated blade record, potential state transition}
blade $\gets$ GetBlade(serialNumber)\;
\If{blade.currentState $\neq$ \textsc{In\_Service}}{
    \Return error: invalid state for flight logging\;
}
blade.totalFlightHours $\gets$ blade.totalFlightHours + flightHours\;
blade.totalFlightCycles $\gets$ blade.totalFlightCycles + flightCycles\;
blade.hoursSinceInspection $\gets$ blade.hoursSinceInspection + flightHours\;
blade.cyclesSinceInspection $\gets$ blade.cyclesSinceInspection + flightCycles\;
blade.daysSinceInspection $\gets$ DaysSince(blade.lastInspectionDate)\;
\BlankLine
\uIf{blade.hoursSinceInspection $\geq$ 500}{
    blade.inspectionDueReason $\gets$ ``HOURS\_EXCEEDED''\;
}
\uElseIf{blade.cyclesSinceInspection $\geq$ 500}{
    blade.inspectionDueReason $\gets$ ``CYCLES\_EXCEEDED''\;
}
\ElseIf{blade.daysSinceInspection $\geq$ 180}{
    blade.inspectionDueReason $\gets$ ``DAYS\_EXCEEDED''\;
}
\BlankLine
\If{blade.inspectionDueReason $\neq$ ``''}{
    blade.currentState $\gets$ \textsc{Inspection\_Due}\;
    EmitEvent(InspectionDueEvent, blade)\;
}
SaveBlade(blade)\;
\end{algorithm}


The algorithm first validates that the blade is in \textsc{In\_Service} state, then updates cumulative and since-inspection counters. If any threshold is exceeded, the state transitions to \textsc{Inspection\_Due} and an \texttt{Inspection\-DueEvent} is emitted for downstream notification. The triggering reason is recorded for audit purposes. The current implementation uses hardcoded thresholds of 500 flight hours, 500 flight cycles, and 180 calendar days; these values are configurable and can be adjusted to match operator-specific maintenance programs.

\paragraph{Inspection Recording}

Algorithm~\ref{alg:inspection} presents the inspection submission logic, which binds AI detection results to on-chain records with cryptographic integrity.

\begin{algorithm}[!htbp]
\caption{SubmitInspectionResult}
\label{alg:inspection}
\KwIn{serialNumber, inspectionData, imageCID, imageHash}
\KwOut{Inspection event recorded, state updated}
blade $\gets$ GetBlade(serialNumber)\;
\If{blade.currentState $\neq$ \textsc{Inspection\_Due}}{
    \Return error: blade not due for inspection\;
}
\BlankLine
mspID $\gets$ GetClientMSPID()\;
inspector $\gets$ GetClientID()\;
\BlankLine
event $\gets$ new InspectionEvent\;
event.eventID $\gets$ serialNumber + ``\_'' + timestamp\;
event.inspector $\gets$ inspector\;
event.organization $\gets$ mspID\;
event.aiModelName $\gets$ inspectionData.modelName\;
event.aiModelVersion $\gets$ inspectionData.modelVersion\;
event.detectedDefects $\gets$ inspectionData.defects\;
event.defectCount $\gets$ Length(inspectionData.defects)\;
event.overallResult $\gets$ inspectionData.result\;
event.imageIPFS $\gets$ imageCID\;
event.imageHash $\gets$ imageHash\;
event.flightHoursAtInspection $\gets$ blade.totalFlightHours\;
event.flightCyclesAtInspection $\gets$ blade.totalFlightCycles\;
\BlankLine
blade.currentState $\gets$ \textsc{Under\_Inspection}\;
blade.inspectionHistory.append(event.eventID)\;
blade.totalInspections $\gets$ blade.totalInspections + 1\;
\BlankLine
\If{event.defectCount $>$ 0}{
    EmitEvent(DefectDetectedEvent, event)\;
}
SaveInspectionEvent(event)\;
SaveBlade(blade)\;
\end{algorithm}


Critically, the inspector identity and organization are derived from the transaction context via \texttt{GetClientMSPID} and \texttt{GetClientID}, ensuring non-repudiable attribution that cannot be falsified by client applications. The inspection event stores the IPFS CID and SHA-256 hash of the inspection image, enabling independent verification of artifact integrity.

\subsection{API Gateway and Off-Chain Services}
\label{subsec:api}

The API gateway, implemented in Node.js with Express, bridges client applications with the Fabric network and off-chain services. It maintains per-organization wallet directories containing enrolled identities, selecting the appropriate wallet based on the \texttt{X-Org} header or endpoint-specific defaults.

The gateway connects to the Fabric network using the \texttt{fabric-network} SDK with connection pooling and automatic reconnection with exponential backoff. For IPFS integration, the gateway communicates with the local Kubo daemon via its HTTP API. The \texttt{/api/ipfs/upload} endpoint accepts image uploads, stores them in IPFS, computes the SHA-256 hash, and returns both the CID and hash for inclusion in inspection submissions.

The traceability proof endpoint (\texttt{/api/blades/:sn/proof}) aggregates the blade record and all inspection events, recomputes SHA-256 hashes of the serialized data, and bundles IPFS retrieval URLs for each inspection image. This proof package can be independently verified by regulators without trusting the API gateway.

Real-time event delivery uses Server-Sent Events (SSE) via the \texttt{/api/\allowbreak events/\allowbreak stream} endpoint, which republishes chaincode events (\texttt{Inspection\-DueEvent}, \texttt{Defect\-Detected\-Event}, \texttt{OEM\-DefectAlert}, \texttt{Blade\-ScrapedEvent}) to subscribed clients.

\subsection{Defect Detection Integration}
\label{subsec:detection}

BladeChain integrates AI-based defect detection as a pluggable, external module rather than embedding inference within the API gateway. This design enables model updates without modifying the core system and supports heterogeneous detection approaches across organizations.

\begin{figure}[!htbp]
    \centering
    \begin{minipage}{0.32\linewidth}
        \centering
        \includegraphics[width=\linewidth]{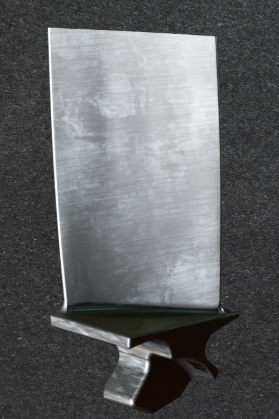}
        \caption*{(a) Healthy blade}
    \end{minipage}
    \hfill
    \begin{minipage}{0.32\linewidth}
        \centering
        \includegraphics[width=\linewidth]{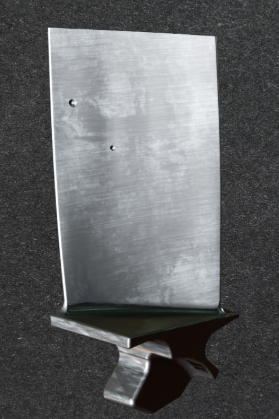}
        \caption*{(b) Defective blade}
    \end{minipage}
    \hfill
    \begin{minipage}{0.32\linewidth}
        \centering
        \includegraphics[width=\linewidth]{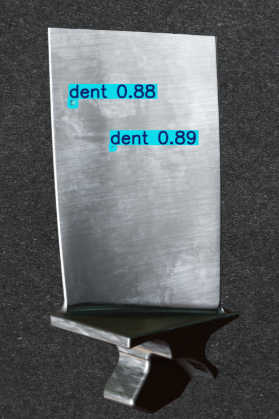}
        \caption*{(c) AI-detection}
    \end{minipage}

    \caption{Blade images at different processing stages: 
(a) healthy blade generated using BladeSynth; 
(b) defective blade generated using BladeSynth; 
(c) AI inference result for the defective image in (b), showing detected defects annotated with bounding boxes.}
    \label{fig:bladesynth}
\end{figure}

The prototype uses a YOLOv8 model trained on the BladeSynth dataset~\cite{BladeSynth2025}, a synthetic dataset of turbine blade images with pixel-level defect annotations. BladeSynth provides controlled generation of healthy and defective blade images, enabling systematic evaluation of detection performance across defect types. Figure~\ref{fig:bladesynth} shows examples of healthy and defective blade images from the dataset.

The detection module accepts a blade image and returns a structured payload containing the model name and version, detected defects with classifications and confidence scores, bounding box coordinates, and an overall result (PASS, FAIL, or CRITICAL based on defect severity). This payload is submitted to the API gateway, which records it on-chain along with the IPFS CID and SHA-256 hash of the original image.

By recording the AI model name and version with each inspection, BladeChain enables auditors to assess detection results in context. If a model is later found to have systematic blind spots, historical inspections using that model can be identified and flagged for re-evaluation, which is a capability absent from systems that treat AI detection as a black box.

\subsection{Client Applications}
\label{subsec:clients}

BladeChain includes a web-based dashboard that provides stakeholder-specific views over the consortium ledger. The application communicates with the API gateway via REST calls and subscribes to real-time events through SSE.

\begin{figure}[!htbp]
\centering
\includegraphics[width=\columnwidth]{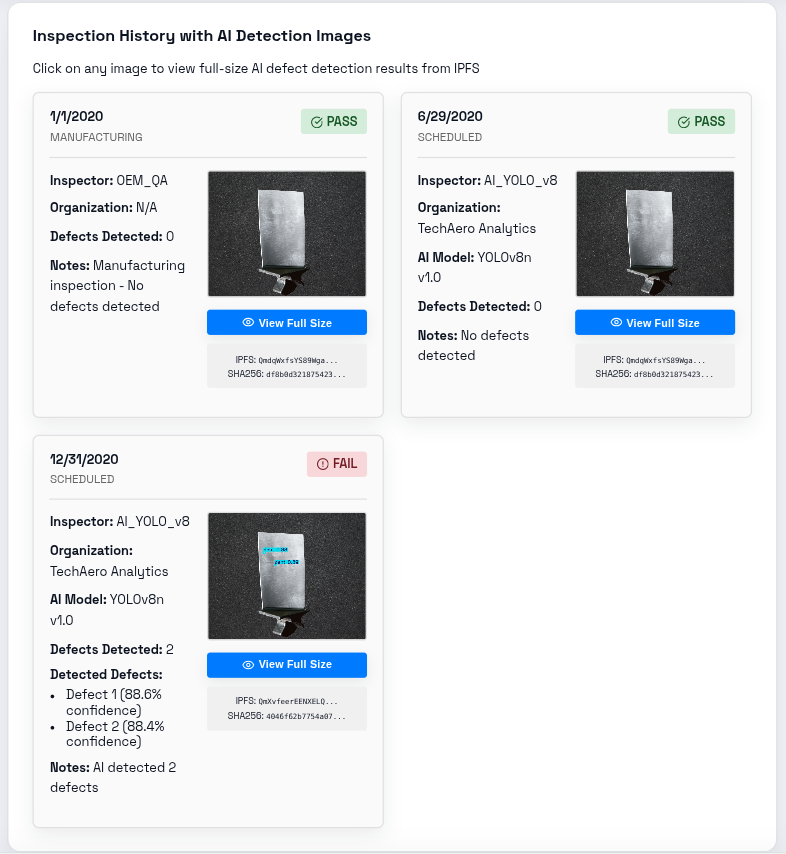}
\caption{AI inspection history view (cropped from blade dossier), showing IPFS-stored images and SHA-256 hashes. The full dossier view also includes blade metrics such as operating hours, cycles, ownership, and current state.}
\label{fig:webapp-history}
\end{figure}

\begin{figure*}[!htbp]
\centering

\includegraphics[width=\textwidth]{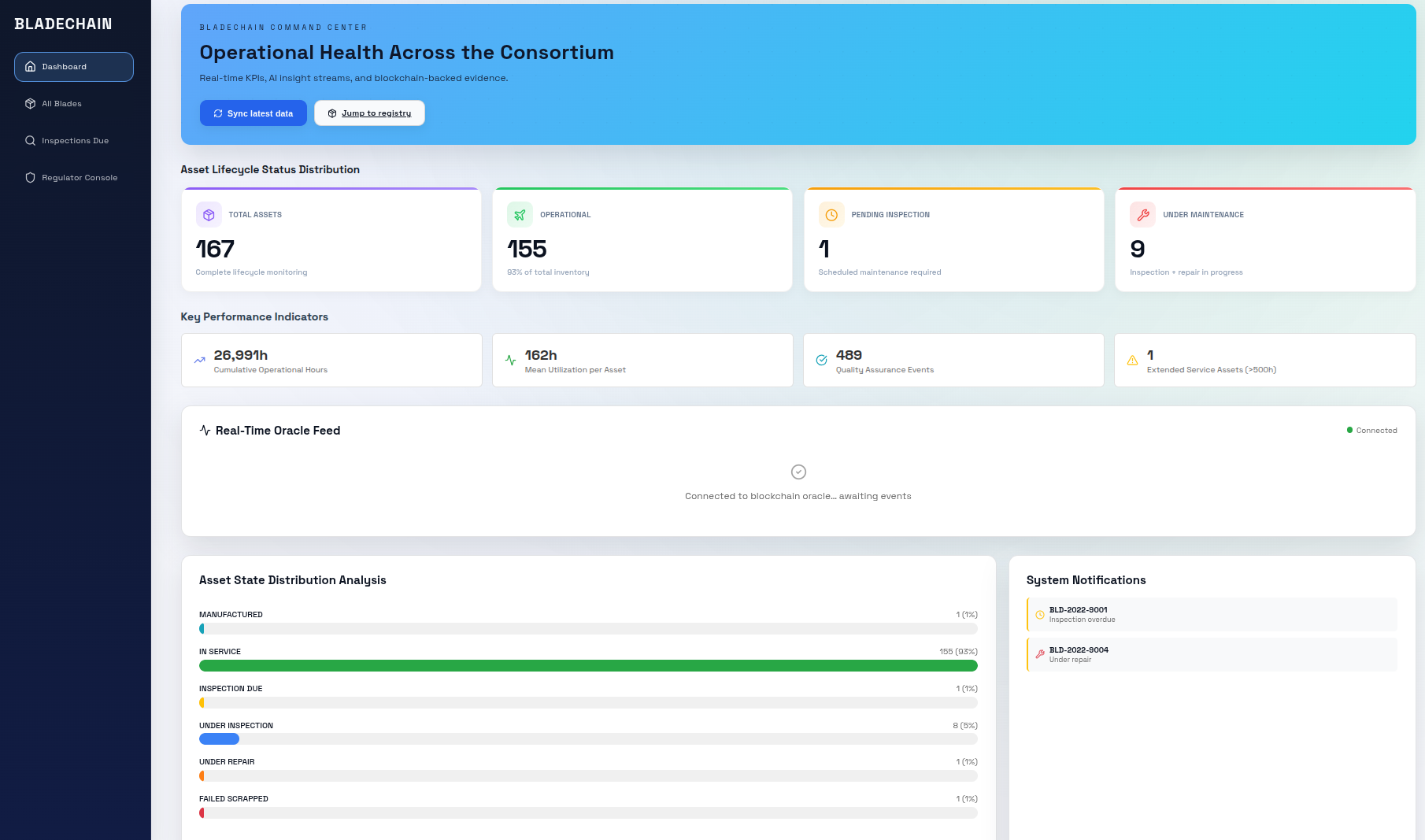}
\caption*{(a) Dashboard overview}

\vspace{1em}

\includegraphics[width=\textwidth]{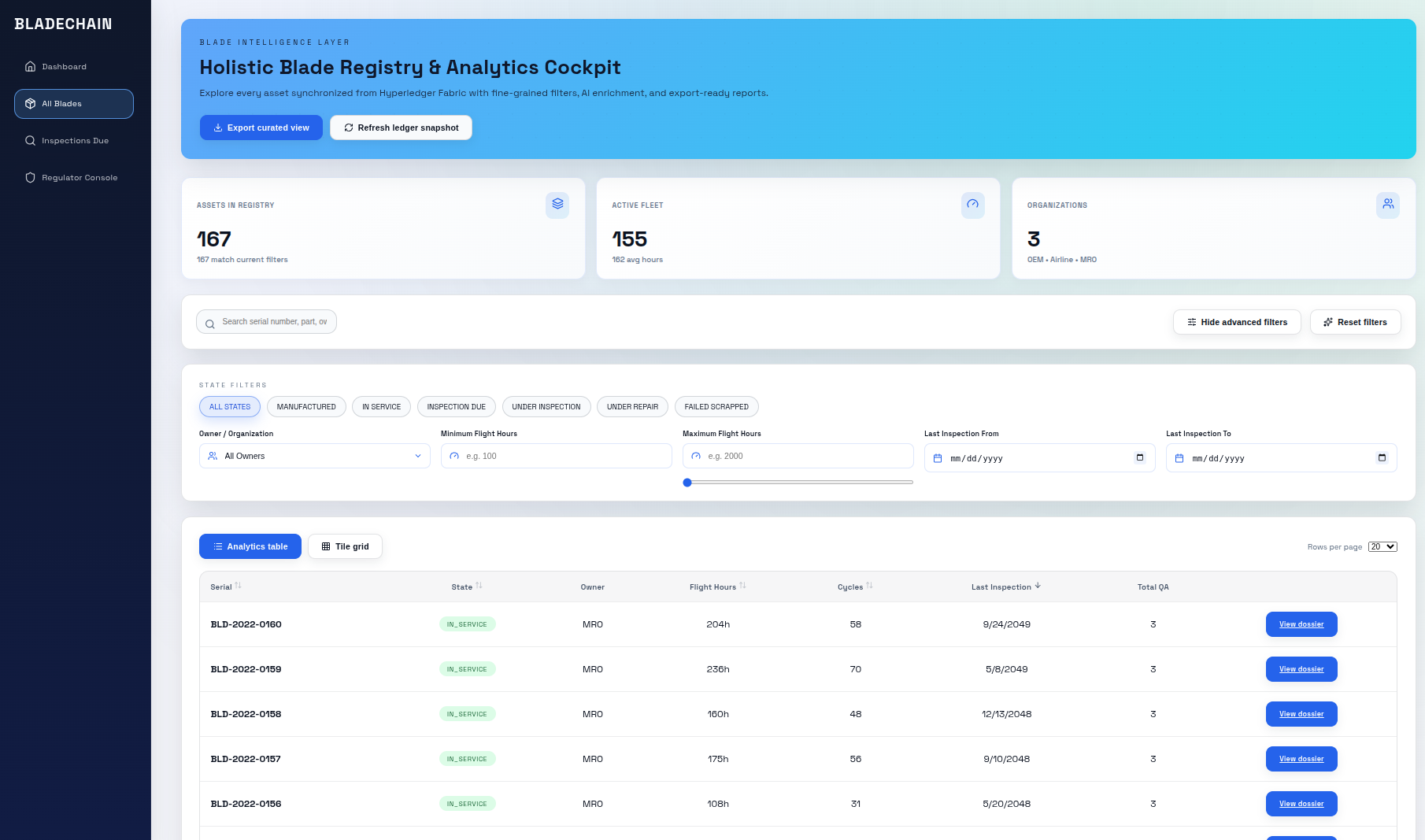}
\caption*{(b) Blade registry with search and filters}

\caption{BladeChain web application interfaces: 
(a) system dashboard summarizing life cycle and integrity metrics; 
(b) blade registry for search, filter, and state tracking.}
\label{fig:webapp-main}
\end{figure*}

Figure~\ref{fig:webapp-main} highlights two core interfaces. The dashboard provides a high-level operational overview, including state distribution, inspection due counts, and storage/AI summary metrics. The blade registry supports search, filtering, and sorting across the entire fleet, allowing users to quickly locate blades by serial number, owner, or life cycle state. The inspection history view, in Figure~\ref{fig:webapp-history} (cropped from a blade dossier page), focuses on AI-assisted inspection events with IPFS-hosted images and cryptographic hashes, while the full dossier page also exposes additional blade metrics such as flight hours, cycles, ownership, and life cycle status. Together, these views provide both system-wide situational awareness and drill-down traceability for individual assets.

%% file: eval.tex
\section{Evaluation}
\label{sec:evaluation}

We evaluate BladeChain through functional validation, performance benchmarking, and security analysis. Our experiments address three questions: (1) Does the system correctly enforce the blade life cycle state machine? (2) How does performance scale with workload size? (3) Does the tamper-evidence mechanism detect artifact manipulation?

\subsection{Experimental Setup}
\label{subsec:setup}

Experiments were conducted on a machine with an Intel Core Ultra 7 165U processor (12 cores, 24 threads), 16~GB RAM, and NVMe SSD storage, running Ubuntu 24.04 LTS with Docker 28.4.0. The Fabric network consisted of four organizations (OEM, Airline, MRO, Regulator), each operating one peer with CouchDB state storage, and a three-node Raft ordering cluster. The \texttt{bladelifecycle} chaincode was deployed on the \texttt{blade-lifecycle-channel} with a majority endorsement policy requiring three of four organizations.

The workload simulator drives blades through complete life cycle scenarios: manufacturing, release to the airline, flight operations (15--35 flights per cycle, 2--8 hours per flight), inspection with AI detection, and repair when defects are found. Each blade completes two full inspection cycles, with inspection thresholds set at 500 flight hours, 500 cycles, or 180 days. These parameters reflect representative airline maintenance scheduling practices, where component inspections are triggered by whichever limit is reached first among flight hours, flight cycles, or calendar time~\cite{saltoglu2016}. The selected thresholds approximate typical A-check maintenance intervals for commercial aircraft~\cite{deng2020}. We evaluated three workload sizes: 10, 50, and 100 blades, generating approximately 430, 2,160, and 4,080 API operations, respectively.

\subsection{Functional Validation}
\label{subsec:functional}

We validated that BladeChain correctly enforces the blade life cycle state machine across all workload sizes. Table~\ref{tab:functional} summarizes the results.

\begin{table}[!htbp] 
\vspace{5pt}
\centering
\caption{Functional validation results.}
\label{tab:functional}
\renewcommand{\arraystretch}{1.3}
\begin{tabular}{lcc}
\toprule
\textbf{Workload} & \textbf{Completed} & \textbf{Operations} \\
\midrule
10 blades $\times$ 2 cycles & 10/10 (100\%) & 429 \\
50 blades $\times$ 2 cycles & 50/50 (100\%) & 2,157 \\
100 blades $\times$ 2 cycles & 100/100 (100\%) & 4,075  \\
\bottomrule
\end{tabular}
\vspace{5pt}
\end{table}

All blades successfully progressed through the complete life cycle: \textsc{Manufactured} $\rightarrow$ \textsc{In\_Service} $\rightarrow$ \textsc{Inspection\_Due} $\rightarrow$ \textsc{Under\_Inspection} $\rightarrow$ \textsc{Under\_Repair} (when defects detected) $\rightarrow$ \textsc{In\_Service}. The chaincode correctly triggered automatic transitions to \textsc{Inspection\_Due} when flight hour, cycle, or calendar thresholds were exceeded. Inspection events were recorded with AI model metadata, IPFS CIDs, and SHA-256 hashes, confirming that evidence binding operates as designed.

\subsection{Performance Evaluation}
\label{subsec:performance}

\paragraph{Throughput and Scalability}

Table~\ref{tab:scalability} presents throughput measurements across workload sizes. The system maintains a consistent throughput of 25.8--26.3 operations per minute regardless of workload size, demonstrating linear scalability. This throughput is appropriate for maintenance workloads where inspections occur every 500+ flight hours rather than in real-time.

\begin{table}[!htbp] 
\vspace{5pt}
\centering
\caption{Scalability evaluation results.}
\label{tab:scalability}
\renewcommand{\arraystretch}{1.5}
\begin{tabular}{lccc}
\toprule
\textbf{Workload} & \textbf{Duration} & \textbf{Throughput} & \textbf{Ops/Blade} \\
\midrule
10 blades & 979 sec & 26.28 ops/min & 42.9 \\
50 blades & 4,933 sec & 26.24 ops/min & 43.1 \\
100 blades & 9,476 sec & 25.80 ops/min & 40.8 \\
\bottomrule
\end{tabular}
\vspace{5pt}
\end{table}

\paragraph{Transaction Latency}

Write operations (state-changing transactions requiring consensus) exhibit a median latency of 2,084~ms with P95 at 2,103~ms. Table~\ref{tab:latency} breaks down latency by operation type for the 100-blade workload. The consistent latency across operation types reflects the uniform cost of Raft consensus and multi-organization endorsement. Read operations, which do not require consensus, complete in 17--22~ms.

\begin{table}[!htbp] 
\centering
\vspace{5pt}
\caption{Transaction latency by operation type.}
\label{tab:latency}
\renewcommand{\arraystretch}{1.1}
\begin{tabular}{lcccc}
\toprule
\textbf{Operation} & \textbf{Count} & \textbf{P50 (ms)} & \textbf{P95 (ms)} & \textbf{Max (ms)} \\
\midrule
Manufacture & 100 & 2,092 & 2,102 & 2,123 \\
Flight log & 1,800+ & 2,085 & 2,105 & 2,107 \\
Inspection & 200+ & 2,079 & 2,095 & 2,098 \\
Repair & 100+ & 2,080 & 2,087 & 2,100 \\
Approval & 100+ & 2,074 & 2,076 & 2,076 \\
\bottomrule
\end{tabular}
\vspace{5pt}
\end{table}

\paragraph{Baseline Comparison}
To contextualize the performance overhead of blockchain-based traceability, we implemented an equivalent system using PostgreSQL 16.11 as a centralized database, deployed locally on the same hardware described in Section~\ref{subsec:setup}. The baseline schema mirrors the chaincode data structures (blades, inspections, flight operations, repairs) with indexes on blade identifiers and state fields. PostgreSQL was configured with default settings, with \texttt{synchronous\_commit=on} and \texttt{fsync=on} to ensure durability semantics comparable to the blockchain's persistence guarantees. Table~\ref{tab:baseline} compares the two approaches.

\begin{table}[!htbp] 
\vspace{5pt}
\centering
\caption{Performance comparison: BladeChain vs. PostgreSQL baseline.}
\label{tab:baseline}
\renewcommand{\arraystretch}{1.1}
\begin{tabular}{lccc}
\toprule
\textbf{Metric} & \textbf{PostgreSQL} & \textbf{BladeChain} & \textbf{Ratio} \\
\midrule
Duration (100 blades) & 92 sec & 9,476 sec & 103$\times$ \\
Throughput & 3,617 ops/min & 25.8 ops/min & 140$\times$ \\
Write latency (avg) & 0.72 ms & 2,087 ms & 2,899$\times$ \\
Read latency (avg) & 0.88 ms & 20.9 ms & 24$\times$ \\
\bottomrule
\end{tabular}
\vspace{5pt}
\end{table}

BladeChain's throughput is approximately 140$\times$ lower than the centralized baseline. This overhead reflects the cost of four-organization consensus, cryptographic endorsement signatures, and append-only ledger semantics. However, these properties, which are impossible to achieve in a centralized database, are precisely what enable multi-party trust without a central authority. For aviation maintenance workloads, where a blade undergoes inspection every 500+ flight hours (months of operation), the two-second transaction latency poses no operational constraint. The tradeoff favors integrity and auditability over raw throughput.

\paragraph{IPFS Overhead}

Inspection images average 2.06~MB each, while on-chain metadata (CID, hash, defect records) requires approximately 0.5~KB, which is a ratio of 0.024\%. IPFS upload latency averages 40~ms and does not significantly impact end-to-end inspection recording time, which is dominated by Fabric consensus overhead.

\subsection{Security Validation}
\label{subsec:security-eval}

\begin{figure*}[!htbp]
  \centering
  \includegraphics[width=\textwidth]{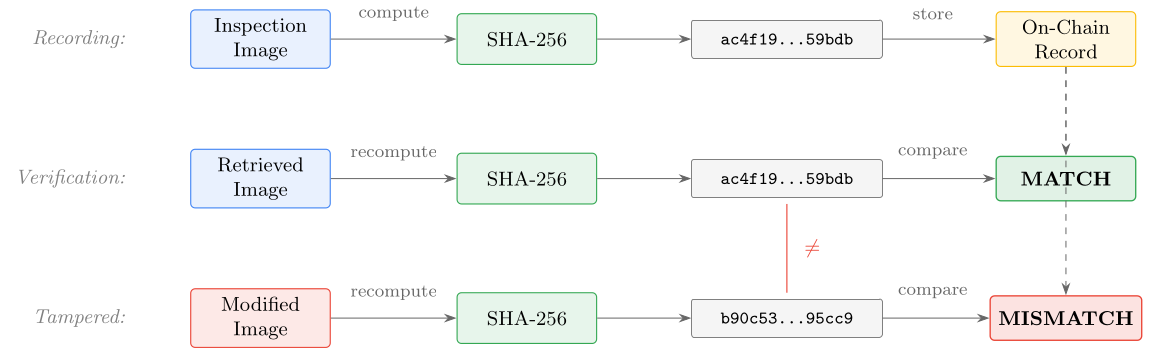}
  \caption{Tamper detection verification. The recomputed hash from the retrieved artifact differs from the on-chain hash, producing a detectable mismatch.}
  \label{fig:tamper}
\end{figure*}

We validated BladeChain's tamper-evidence mechanisms through controlled experiments.

\paragraph{IPFS Hash Integrity}

Figure~\ref{fig:tamper} illustrates the tamper detection mechanism. At inspection time, the SHA-256 hash of the image is computed and stored on-chain alongside the IPFS CID. To verify integrity, the image is retrieved from IPFS, and its hash is recomputed. If the hashes match, the artifact is authentic; any modification produces a different hash, resulting in a detectable mismatch. 
In our test runs, we uploaded an inspection image, recorded its hash (\texttt{ac4f193f...}), modified the image bytes, and recomputed the hash (\texttt{b90c534b...}); the system flagged the mismatch in all trials as expected. Hash recomputation averaged 17.95~ms (max 22.34~ms), showing that verification remains lightweight.

\paragraph{Blockchain Immutability}

We verified that the append-only ledger prevents historical modification by attempting to alter previously committed blade records. The Fabric ledger correctly rejected all modification attempts, and the complete history remained intact and verifiable through the \texttt{GetBladeCompleteHistory} query.

\begin{figure}[!htbp]
\vspace{10pt}
  \centering
  \includegraphics[width=\columnwidth]{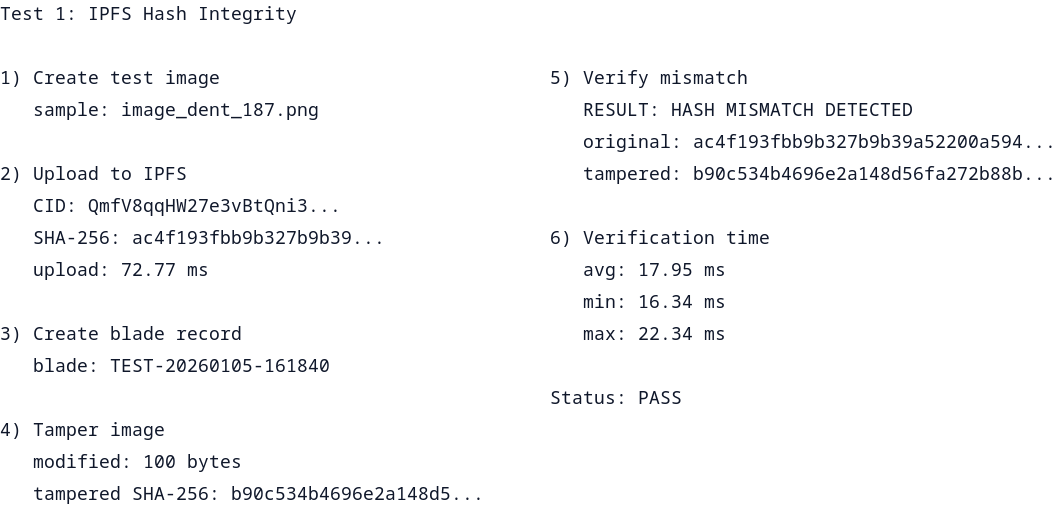}
  \caption{Console output excerpt from the tamper-evidence test, confirming the hash mismatch and verification timing.}
  \label{fig:security}
  \vspace{5pt}
\end{figure}

Figure~\ref{fig:security} provides the corresponding console evidence for the security validation results.

These results confirm that BladeChain provides the tamper-evidence guarantees required for regulatory compliance: inspection artifacts cannot be modified without detection, and historical records remain immutable once committed to the distributed ledger.

%% file: conc.tex
\section{Conclusion}
\label{sec:conclusion}

This paper presented BladeChain, a blockchain-based framework for aircraft engine blade inspection traceability. By combining Hyperledger Fabric's permissioned ledger with IPFS-based artifact storage and pluggable AI defect detection, BladeChain addresses critical gaps in existing aviation maintenance systems: the lack of multi-party trust, the absence of tamper-evident evidence binding, and the inability to audit AI-assisted inspection decisions.

Our design enforces inspection schedules through chaincode-level state machine transitions, eliminating the manual tracking errors that plague current systems. The multi-organization endorsement policy ensures that no single party can unilaterally create or modify inspection records, while cryptographic binding of inspection images via SHA-256 hashes and IPFS CIDs enables verification of artifact integrity. By recording the AI model name and version for each inspection, BladeChain provides the provenance information regulators need to assess detection results and identify blades inspected by models later found to be deficient.

Evaluation of workloads up to 100 blades demonstrated 100\% functional correctness across the complete life cycle state machine, with consistent throughput of approximately 26 operations per minute. While this represents a reduction compared to a centralized PostgreSQL baseline, the overhead reflects the cost of four-organization consensus and immutable ledger semantics, which are properties essential for multi-party trust and regulatory compliance. Security validation confirmed that artifact tampering is detected within 17~ms through hash verification, and the append-only ledger prevents unauthorized modification of historical records.

BladeChain demonstrates that blockchain technology can provide the trust, transparency, and auditability required for safety-critical aviation maintenance workflows. As regulatory frameworks increasingly mandate digital traceability, systems like BladeChain offer a path toward tamper-evident maintenance records that all stakeholders can independently verify and trust~\cite{Risso2023}.

Future work will explore integration with legacy maintenance management systems such as AMOS and TRAX, positioning BladeChain as an automated verification layer rather than a replacement for existing workflows. This approach would allow stakeholders to retain their current software investments while gaining the benefits of blockchain-based traceability, with BladeChain operating either as a standalone solution with its web dashboard or as a backend verification service integrated via standard APIs.

%% file: ack.tex
\section{Acknowledgments}




This research was funded by Khalifa University of Science and Technology through the Faculty Start-Up (FSU) Fund under Project ID: KU-INT-FSU-2024-8474000660. 
